\documentclass[prc,twocolumn,tightenlines,showpacs,floatfix]{revtex4}
\usepackage{dcolumn}
\usepackage{bm}
\usepackage{graphicx}
\begin{document}

{
\title{Nucleon and nucleon-pair momentum distributions in $A\leq 12$ nuclei}
\author{R. B. Wiringa$^1$}
\email{wiringa@anl.gov}
\author{R. Schiavilla$^{2,3}$}
\email{schiavil@jlab.org}
\author{\mbox{Steven C. Pieper$^1$}}
\email{spieper@anl.gov}
\author{J. Carlson$^4$}
\email{carlson@lanl.gov}
\affiliation{
$^1$Physics Division, Argonne National Laboratory, Argonne, IL 60439\\
$^2$Theory Center, Jefferson Laboratory, Newport News, VA 23606 \\
$^3$\mbox{Department of Physics, Old Dominion University, Norfolk, VA 23529}\\
$^4$\mbox{Theoretical Division, Los Alamos National Laboratory, Los Alamos, NM 87545}
}

\date{\today}

\begin{abstract}
We report variational Monte Carlo calculations of single-nucleon momentum 
distributions and nucleon-pair and nucleon-cluster momentum distributions 
for $A \leq 12$ nuclei.
The wave functions have been generated for a Hamiltonian containing the Argonne
v$_{18}$ two-nucleon and Urbana X three-nucleon potentials.
The single-nucleon and nucleon-pair momentum distributions exhibit universal
features attributable to the pion-exchange tensor interaction.
The single-nucleon distributions are broken down into proton and neutron 
components and spin-up and spin-down components where appropriate.
The nucleon-pair momentum distributions are given either in pair spin and
isospin $ST$ projection or for $pp$, $pn$, and $nn$ pairs.
The nucleon-cluster momentum distributions include $dp$ in $^3$He,
$tp$ and $dd$ in $^4$He, $\alpha d$ in $^6$Li, $\alpha t$ in $^7$Li, 
and $\alpha\alpha$ in $^8$Be.
Detailed tables are provided online for download.
\end{abstract}


\maketitle

}

\section {Introduction}
\label{sec:intro}

Momentum distributions of nucleons and nucleon clusters in nuclei provide
useful insights into various reactions on nuclei, such as
$(e,e^\prime p)$ and $(e,e^\prime pp/pn)$ electrodisintegration processes
or neutrino-nucleus interaction experiments.
Variational Monte Carlo (VMC) calculations of these distributions
in $A$=3,4 nuclei were first reported in~\cite{SPW86}.
They were subsequently improved upon and extended to 
light $p$-shell nuclei, and also to the study of nucleon-pair momentum
distributions~\cite{SWPC07,WSPC08}.
Many results of these studies have been supplied on request to various
researchers, but not formally published before now.
Hence, in this paper, we detail their current status, provide
figures illustrating the features of these momentum distributions,
and announce a web site with detailed tabulations available for
download~\cite{momenta}.

Specifically, we present single-nucleon momentum distributions for
$A$=2--12 nuclei, and $d$, $t$, and $\alpha$-cluster distributions for
$A \leq 8$ nuclei.
We also provide a number of nucleon-pair momentum distributions as a 
function of either the relative pair momentum $q$ or the pair center-of-mass
momentum $Q$. 
These are projected into spin $S=0$ or 1 and isospin $T=0$ or 1
combinations, or alternatively into $pp$, $pn$, and $nn$ pairs.
Finally we provide more complete calculations of nucleon-pair distributions
as a function of both $q$ and $Q$ in $A$=3,4 nuclei than previously available.

This work is complementary to a major recent study of nucleon momentum 
distributions by Alvioli {\it et al.}~\cite{Alvioli13}.
They made {\it ab initio} calculations of the $A$=2,3,4 $s$-shell nuclei
using a variety of realistic two-nucleon interactions, as well as more
approximate calculations of the larger closed-shell nuclei $^{16}$O and 
$^{40}$Ca, with an emphasis on the spin-isospin origin of the distributions.
Our work fills in the light $p$-shell nuclei up to $^{12}$C, using primarily
(for reasons discussed below) a realistic two- plus three-nucleon interaction.
Where our calculations overlap, we are in substantial quantitative agreement 
with their results.

The VMC wave functions used in our calculations have been obtained 
for a realistic Hamiltonian consisting of nonrelativistic nucleon kinetic
energy, the Argonne $v_{18}$ (AV18) two-nucleon ($N\!N$)~\cite{WSS95},
and Urbana X (UX) three-nucleon ($3N$) interactions.
Urbana X is intermediate between the Urbana IX (UIX) and Illinois-7 (IL7)
potentials~\cite{PPWC01}. 
It has the long-range two-pion $P$-wave term and short-range central repulsion 
of UIX supplemented with a two-pion $S$-wave term, with the strengths of these 
three terms taken from IL7~\cite{P08}; it does not have the three-pion-ring 
or isospin-dependent short-range repulsive terms of IL7. 

The VMC wave functions have the general form
\begin{equation}
\label{eq:psiv}
   |\Psi_V\rangle =
      {\cal S} \prod_{i<j}^A
      \left[1 + U_{ij} + \sum_{k\neq i,j}^{A}\tilde{U}^{TNI}_{ijk} \right]
      |\Psi_J\rangle \ ,
\end{equation}
where ${\cal S}$ is a symmetrization operator and $U_{ij}$ and 
$\tilde{U}^{TNI}_{ijk}$ are two- and three-body correlation operators.
The single-particle Jastrow wave function $\Psi_J$ is fully antisymmetric 
and has the $(J^\pi;T,T_z)$ quantum numbers of the state of interest.
This construction has been described in detail elsewhere~\cite{W91,PPCPW97} 
and the variational parameters in $\Psi_V$ have been optimized by minimizing
the variational energy expectation value, subject to the constraint of
obtaining approximately correct charge radii.
Most of these wave functions have been used as starting trial functions for 
recent calculations of energies, electromagnetic moments and transitions,
and spectroscopic overlaps using the more accurate Green's function Monte Carlo 
(GFMC) method~\cite{BPW11,M+PW12,PPSW13,D+PW13}.

The probability of finding a nucleon with momentum $k$ and spin-isospin 
projection $\sigma$,$\tau$ in a given nuclear state 
is proportional to the density
\begin{eqnarray}
\label{eq:rhok}
\rho_{\sigma\tau}({\bf k})\!\!&=&\!\!
\int d{\bf r}^\prime_1\, d{\bf r}_1\, d{\bf r}_2 \cdots d{\bf r}_A\,
\psi^\dagger_{JM_J}({\bf r}_1^\prime,{\bf r}_2, \dots,{\bf r}_A)\,  \nonumber \\
& \times &\, e^{-i{\bf k}\cdot ({\bf r}_1-{\bf r}^\prime_1)}
\, P_{\sigma\tau}(1) \, 
\psi_{JM_J} ({\bf r}_1,{\bf r}_2, \dots,{\bf r}_A) \, .
\end{eqnarray}
$P_{\sigma\tau}(i)$ is the spin-isospin projection operator for nucleon $i$,
and $\psi_{JM_J}$ is the nuclear wave function with total spin $J$ and 
spin projection $M_J$.  The normalization is
\begin{equation}
\label{eq:nst}
  N_{\sigma \tau} = 
    \int \frac{d{\bf k}}{(2\pi)^3}\,\,\rho_{\sigma\tau}({\bf k}) \ ,
\end{equation}
where $N_{\sigma \tau}$ is the number of spin-up or spin-down protons
or neutrons.

The Fourier transform in Eq.~(\ref{eq:rhok}) is computed by Monte
Carlo (MC) integration.  A standard Metropolis walk, guided by 
$|\psi_{JM_J} ({\bf r}_1,\dots,{\bf r}_i,\dots,{\bf r}_A)|^2$, is used
to sample configurations~\cite{PPCPW97}.  
We average over all particles $i$ in each configuration, and for each
particle, a grid of Gauss-Legendre points ${\bf x}_i$ is used to compute 
the Fourier transform.
Instead of just moving the position ${\bf r}_i^\prime$ in the left-hand
wave function away from a fixed position ${\bf r}_i$ in the right-hand 
wave function, both positions are moved 
symmetrically away from ${\bf r}_i$, so Eq.~(\ref{eq:rhok}) becomes
\begin{eqnarray}
\label{eq:actual}
  \rho_{\sigma\tau}({\bf k}) &=& \frac{1}{A}\sum_i
  \int d{\bf r}_1 \cdots d{\bf r}_i \cdots d{\bf r}_A \int d{\bf \Omega}_x
  \int_0^{x_{\rm max}} x^2 dx \nonumber \\
  & & \psi^\dagger_{JM_J} 
  ({\bf r}_1,\dots,{\bf r}_i+{\bf x}/2,\dots,{\bf r}_A) \,
  e^{-i{\bf k}\cdot {\bf x} }  \\
  &\times& \,  P_{\sigma\tau}(i) \,
  \psi_{JM_J} 
  ({\bf r}_1,\dots,{\bf r}_i-{\bf x}/2,\dots,{\bf r}_A) \nonumber \, .
\end{eqnarray}
Here the polar angle $d{\bf \Omega}_x$ is also sampled by MC integration, with 
a randomly chosen direction for each particle in each MC configuration.
This procedure is similar to that adopted in studies of the nucleon-pair
momentum distribution~\cite{SWPC07} and has the advantage of very substantially
reducing the statistical errors originating from the rapidly oscillating
nature of the integrand for large values of $k$.  
To reach momenta $k \sim 10$ fm$^{-1}$ in $^4$He with good statistics 
requires integrating to $x_{\rm max}$=20 fm using 200 Gauss-Legendre points.

\section {Single-nucleon and nucleon-cluster results}
\label{sec:single}

\begin{figure}[t!]
\includegraphics[angle=-90,width=3.25in]{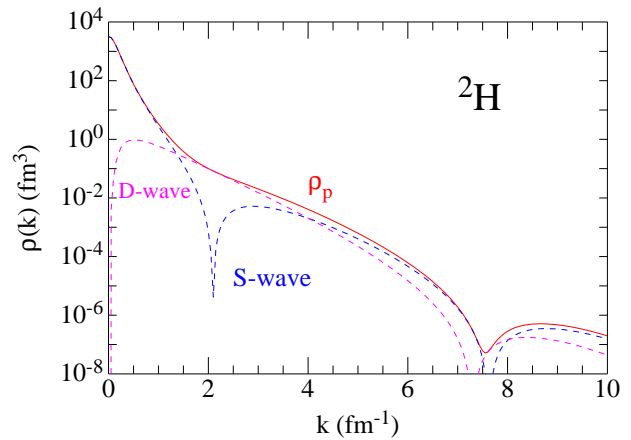}
\caption{(Color online) The total proton momentum distribution in the
deuteron is shown by the red solid line; the contribution from $S$-wave
and $D$-wave components are shown separately by blue and magenta dashed lines.}
\label{fig:deut}
\end{figure}
 
The proton momentum distribution $\rho_p(k)$ in the deuteron is shown in 
Fig.~\ref{fig:deut}.
(For $T_z$=0, our wave functions have $\rho_n$=$\rho_p$.)
In this case $\rho_p(k)$ has been evaluated by direct numerical solution
of the Schr{\"o}dinger equation, although we have checked that our MC code 
gives the same results within statistical errors.
The separate contributions of the $S$- and $D$-wave components of the
deuteron wave function are also shown.
The $S$-wave momentum density has a prominent node at 2 fm$^{-1}$, but this 
is filled in by the $D$-wave momentum density, so the total $\rho_p(k)$ has
a distinctive change of slope at 1.5 fm$^{-1}$, followed by a
broad shoulder out to 7 fm$^{-1}$ before the first $D$-wave
and second $S$-wave nodes occur.
The $D$-wave component is due to the pion-exchange tensor force in AV18.
The broad shoulder is the dominant feature in all the single-nucleon
momentum distributions of larger nuclei shown below.

\begin{figure}[b!]
\includegraphics[angle=-90,width=3.25in]{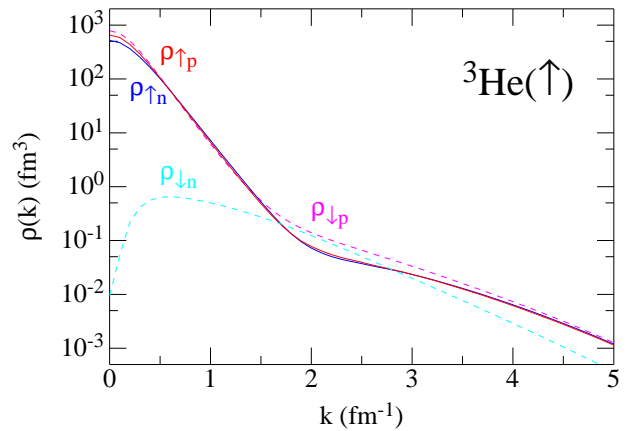}
\caption{(Color online) The spin-isospin densities for polarized 
$^3$He in the $M_J = + \frac{1}{2}$ state are shown
by solid red (blue) lines for spin-up protons (neutrons) and by
dashed magenta (cyan) lines for spin-down protons (neutrons).}
\label{fig:he3pol}
\end{figure}

The polarized proton and neutron densities in $^3$He in the 
$M_J = +\frac{1}{2}$ state are shown in Fig.~\ref{fig:he3pol}.
The spin-up proton and neutron densities are very similar.
The spin-down proton density is slightly larger, particularly
in the dip region around 2 fm$^{-1}$. 
Although the spin-down neutron density is non-zero only by virtue of the tensor 
force, it exceeds the spin-up neutron density in the dip region.
The total normalizations for $^3$He are given in Table~\ref{tab:sigma-tau},
along with those of other nuclei in our study.
The total number of spin-down neutrons is 2/3 of the $^3$He $D$-state 
percentage.
These momentum distributions may be instructive when considering the
use of polarized $^3$He as a polarized neutron target.

\begin{table}[t!]
\caption{Total number of spin-up/down and proton/neutron nucleons in
$J > 0$ nuclei with $M_J = J$.}
\begin{ruledtabular}
\begin{tabular}{ l d d d d }
Nucleus & \multicolumn{1}{c}{$N_{\uparrow p}$} 
        & \multicolumn{1}{c}{$N_{\downarrow p}$}
        & \multicolumn{1}{c}{$N_{\uparrow n}$}
        & \multicolumn{1}{c}{$N_{\downarrow n}$} \\
\hline
$^3$He($\frac{1}{2}^+$)  & 0.974 & 1.026 & 0.938 & 0.062 \\
$^6$Li($1^+$)            & 1.924 & 1.076 & 1.924 & 1.076 \\
$^7$Li($\frac{3}{2}^-$)  & 1.934 & 1.066 & 1.981 & 2.019 \\
$^8$Li($2^+$)            & 1.914 & 1.086 & 2.855 & 2.145 \\
$^9$Li($\frac{3}{2}^-$)  & 1.907 & 1.093 & 3.084 & 2.916 \\
$^9$Be($\frac{3}{2}^-$)  & 1.994 & 2.006 & 2.880 & 2.120 \\
$^{10}$B($3^+$)          & 2.901 & 2.099 & 2.901 & 2.099 
\label{tab:sigma-tau}
\end{tabular}
\end{ruledtabular}
\end{table}
 
\begin{figure}[t!]
\includegraphics[angle=-90,width=3.25in]{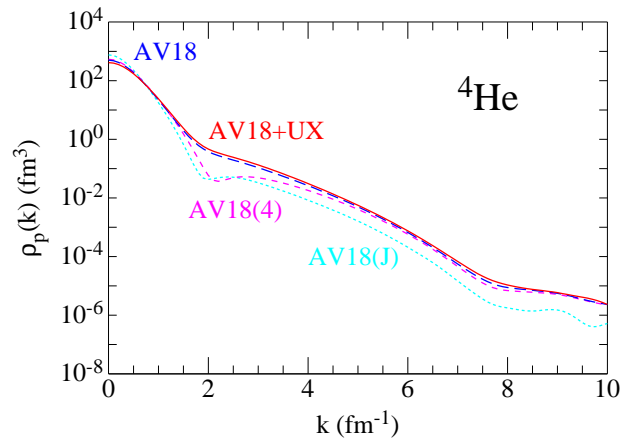}
\caption{(Color online) The proton momentum distribution in $^4$He
is shown for the AV18+UX Hamiltonian by the red solid line, and for
AV18 alone by the blue long-dash line; simplified calculations for
AV18 using 1) only a central Jastrow correlation (J) is shown by the
cyan short-dash line, and 2) including spin-isospin (but no tensor)
correlations (4) is indicated by the magenta dash line.}
\label{fig:he4}
\end{figure}

The proton momentum density in $^4$He is shown in Fig.~\ref{fig:he4}
out to 10 fm$^{-1}$ on the same scale as the deuteron $\rho_p(k)$ in
Fig.~\ref{fig:deut}.
The overall shape is rather similar to the deuteron, with a change of slope
near 2 fm$^{-1}$ and a broad shoulder out to a second dip near 8 fm$^{-1}$.
The $\rho_p(k)$ is shown for the standard AV18+UX Hamiltonian used in
this work, and for AV18 alone using three different versions of the
VMC wave function of Eq.(\ref{eq:psiv}).

The full AV18+UX Hamiltonian gives a VMC energy of $-27.6$ MeV
($-28.3$ MeV in GFMC) and a radius of 1.44 fm, which are close to the
experimental energy of $-28.3$ MeV and point proton radius of 1.46 fm.
The AV18 two-nucleon potential alone gives less binding with a VMC
energy of $-23.7$ MeV ($-24.1$ MeV in GFMC) and a larger point proton 
radius $r_p$ of 1.52 fm.
Because AV18+UX produces energies and radii that are closer to experiment 
for all $A \geq 3$ nuclei, this is the primary model we use for this paper.
The main role of the three-nucleon potential is to fix both the binding
energy and size of the nuclei, which for the $\rho_p(k)$ translates into
an overall shift of the momentum density to larger $k$, with a reduced value 
for $k \leq 1$ fm$^{-1}$.
The $^4$He $D$-state increases from 13\% with AV18 alone to 15\%, and the 
$\rho_p(k)$ beyond 2 fm$^{-1}$ increases by 10-20\%.

To study the source of the intermediate- and high-momentum
densities, we show two AV18 calculations with simplified wave functions:
1) a pure central Jastrow wave function labeled AV18(J) in the figure, 
and 2) a wave function including spin-isospin operator correlations,
but no tensor components, labeled AV18(4).
The AV18(J) calculation has just the $|\Psi_J\rangle$ in Eq.(\ref{eq:psiv}).
The AV18(4) adds to the central Jastrow term the $U_{ij}$ pair correlation 
operator with spin, isospin, and spin-isospin components.
The full AV18 wave function also includes tensor and tensor-isospin components
in the $U_{ij}$ operator.
(For AV18 alone, there is no $\tilde{U}^{TNI}_{ijk}$ correlation.)

The central Jastrow term dominates the momentum density for $k \leq 1$ 
fm$^{-1}$, goes through a minimum at $k=2$ fm$^{-1}$, and then gives
about 30\% of the total density for $k > 3$ fm$^{-1}$.
The spin-isospin correlations in the AV18(4) calculation shift the
Jastrow result to slightly higher momenta, and then begin to dominate at
higher momenta, providing an additional 40-60\% of the density beyond
$k = 4$ fm$^{-1}$.
The tensor components in the full AV18 wave function fill in the minimum in the
region $1.5 \leq k \leq 3$ fm$^{-1}$, just as they do in the deuteron,
but are then only 10-20\% of the total at higher momenta.

\begin{figure}[b!]
\includegraphics[angle=-90,width=3.25in]{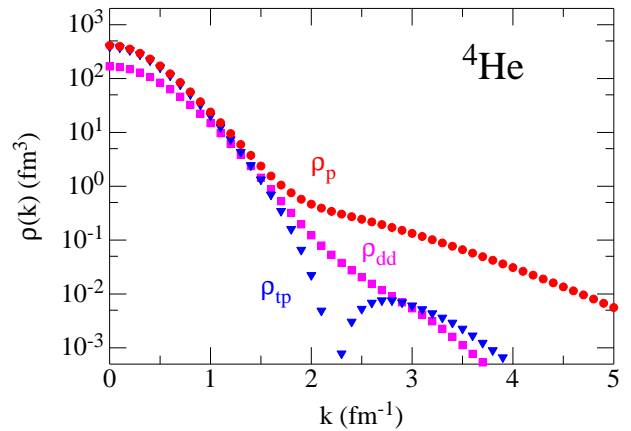}
\caption{(Color online) The proton momentum distribution in $^4$He
is shown by the red circles; the $tp$ cluster distribution is
shown by the blue triangles and the $dd$ cluster distribution is
shown by the magenta squares.}
\label{fig:he4b}
\end{figure}

The proton momentum distribution for $^4$He is also shown in closeup in
Fig.~\ref{fig:he4b} along with the nucleon-cluster distributions 
for $tp$ and $dd$ overlaps.
The definition for these distributions is given in Ref.~\cite{SPW86};
essentially, they are the Fourier transforms of spectroscopic overlaps
as discussed in Ref.~\cite{BPW11}.
The $\rho_{tp}(k)$ at low momenta is almost on top of $\rho_p(k)$,
indicating that in this region, most of the residual nucleus is in
the triton ground state.
The total $N_{tp} = 1.62$, compared to $N_p = 2$, in good agreement with the
spectroscopic factor determined in GFMC calculations~\cite{BPW11}.
The $\rho_{dd}(k)$ integrates to $N_{dd} = 1.005$, with 0.98 coming from
having the $dd$ pair in a relative $S$-wave, and 0.025 from a relative $D$-wave.
Because there are two deuterons in a $dd$ pair, this means there are 
two deuterons among the six pairs in $^4$He from this
configuration. 
(Approximately 0.4 more deuterons should be present in $^4$He from 
$d+pn$ configurations where the $pn$ pair is in a $T$=0 state orthogonal to the 
deuteron~\cite{SPW86}.)

\begin{figure}[t!]
\includegraphics[angle=-90,width=3.25in]{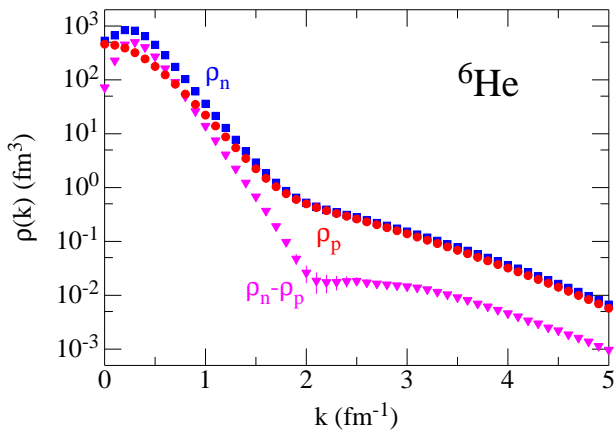}
\caption{(Color online) The proton (neutron) momentum distribution in $^6$He
is shown by the red circles (blue squares); the difference of neutron and
proton densities is shown by the magenta triangles with error bars.}
\label{fig:he6}
\end{figure}

\begin{figure}[b!]
\includegraphics[angle=-90,width=3.25in]{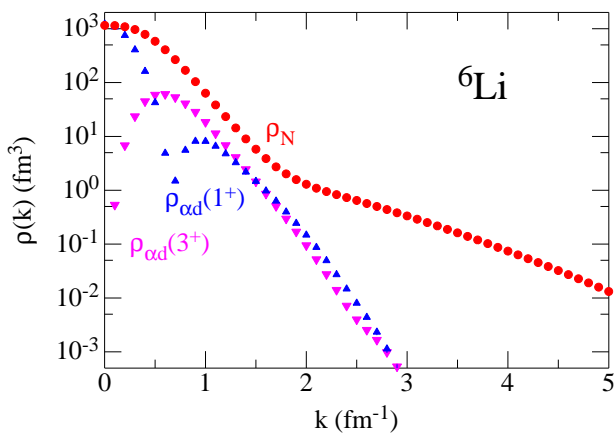}
\caption{(Color online) The nucleon momentum distribution 
$\rho_N$=$\rho_p$+$\rho_n$ in $^6$Li is shown by the red circles; 
the $\alpha d$ cluster distribution for the ground state is shown by blue 
point-up triangles and for the first excited state by magenta point-down
triangles.}
\label{fig:li6}
\end{figure}

The proton and neutron momentum distributions in $^6$He are shown in
Fig.~\ref{fig:he6}.
The proton momentum distribution is close to that in $^4$He.
The two neutrons in the $p$-shell make the neutron momentum distribution
peak at finite $k \sim 0.25$ fm$^{-1}$.
Also shown is the difference $\rho_n(k)-\rho_p(k)$, which should give
a good approximation to the distribution of the $p$-shell halo neutrons.
The difference has a dip at zero momentum, appropriate for the $p$-shell,
and because the halo neutrons are primarily in a relative $^1S_0$ pair,
there is a dip at the usual $S$-wave node position.

Although there are twice as many neutrons as protons in $^6$He, the
high-momentum densities are almost equal; for $q \geq 2$ fm$^{-1}$ the
ratio $\rho_n$/$\rho_p$ = 1.1.
A similar situation holds for $^8$He (not shown here, but illustrated
and tabulated online~\cite{momenta}) where there are three times as many 
neutrons as protons, but above 2 fm$^{-1}$ the ratio $\rho_n$/$\rho_p$ = 1.2.
This supports the recent suggestion of Sargsian~\cite{Sargsian13} that
in neutron-rich systems, the fraction of protons at high momenta is
substantially larger than the fraction of neutrons. 
%which may have 
%implications for the EMC effect in large asymmetric nuclei and for 
%proton properties in neutron stars.

The $^6$Li $J^\pi \!=\! 1^+$ ground state single-nucleon momentum distribution and the
$\alpha d$ cluster distributions in the ground state and first $3^+$ excited 
state are shown in Fig.~\ref{fig:li6}.
In this case we plot $\rho_N$=$\rho_p$+$\rho_n$=$2\rho_p$ for comparison
to the ground state $\alpha d$ distribution to illustrate that at zero momentum
the two are almost equal.
The $\rho_{\alpha d}(k)$ cluster distribution in the $1^+$ ground state has a node 
at $k \sim 0.7$ fm$^{-1}$ because the $\alpha$ and $d$ are in a relative $1s$
spatial state due to antisymmetry~\cite{FPPWSA96}.
The integrated $N_{\alpha d} = 0.86$ is a sum of $S$- and $D$-wave parts of
0.846 and 0.017, respectively.
The $\rho_{\alpha d}(k)$ in the $3^+$ state has the $\alpha$ and $d$ 
in a relative $0d$ spatial state and a somewhat greater $N_{\alpha d} = 0.94$.
However, $\rho_N(k)$ for the excited state would be
indistinguishable from the ground state in this figure, so it is not shown
here, but it is tabulated online.

\begin{figure}[t!]
\includegraphics[angle=-90,width=3.25in]{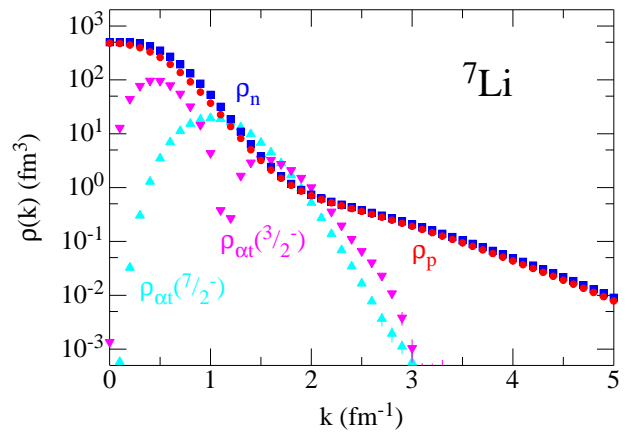}
\caption{(Color online) The proton (neutron) momentum distributions in 
$^7$Li are shown by the red circles (blue squares);
the $\alpha t$ cluster distribution for the ground (second excited) state 
is shown by magenta point-down (cyan point-up) triangles.}
\label{fig:li7}
\end{figure}

The ground state single-nucleon momentum distributions in $^7$Li are shown in 
Fig.~\ref{fig:li7}. 
The proton momentum distribution is similar to that in $^6$Li; the neutron
distribution is bigger at small finite momentum due to the extra $p$-shell
neutron.
The $\alpha t$ cluster distribution is shown for both the $\frac{3}{2}^-$
ground state and the $\frac{7}{2}^-$ second excited state.
The former has a node at zero momentum and a second node at $k \sim 1.1$ 
fm$^{-1}$, indicative of the $\alpha$ and $t$ clusters being in a relative 
$1p$ spatial state, with $N_{\alpha t} = 1.00$.
The $\frac{1}{2}^-$ first excited state has a similar $\rho_{\alpha t}(k)$
shape and is not shown.
The $\frac{7}{2}^-$ second excited state has the $\alpha$ and $t$ clusters
in a relative $0f$ spatial state; the $\frac{5}{2}^-$ third excited state
(not shown) has a similar cluster structure.
The single-nucleon and $\alpha t$ cluster distributions for all four
states are given online~\cite{momenta}.

\begin{figure}[t!]
\includegraphics[angle=-90,width=3.25in]{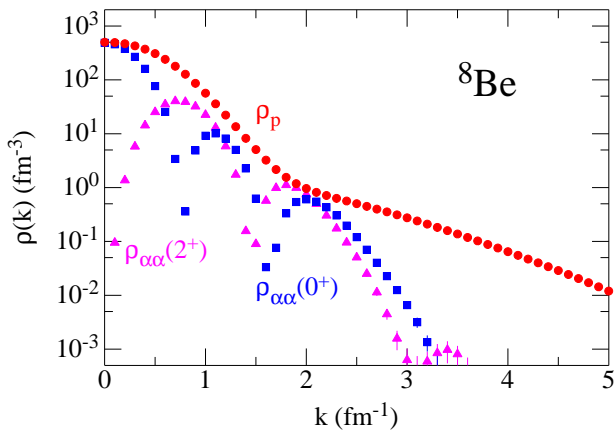}
\caption{(Color online) The proton momentum distribution in $^8$Be is shown 
by the red circles; the $\alpha\alpha$ cluster distribution in the $0^+$ ground
state is shown by the blue squares and in the $2^+$ first excited state 
by the magenta triangles.}
\label{fig:be8}
\end{figure}

\begin{figure}[b!]
\includegraphics[angle=-90,width=3.25in]{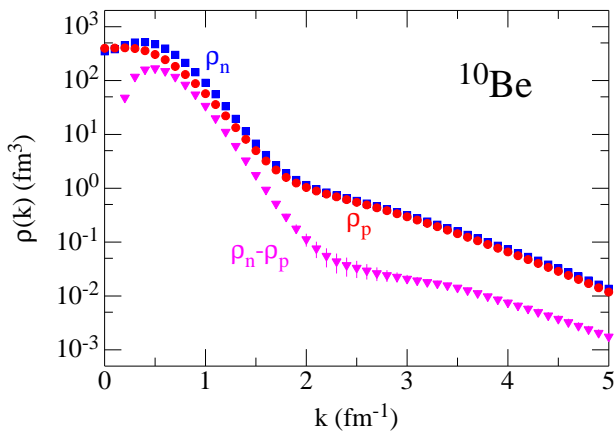}
\caption{(Color online) The proton (neutron) momentum distribution in $^{10}$Be
is shown by the red circles (blue squares); the difference of neutron and
proton densities is shown by the magenta triangles with error bars.}
\label{fig:be10}
\end{figure}

The proton momentum distribution for the $0^+$ ground state of $^8$Be is 
shown in Fig.~\ref{fig:be8} along with the $\alpha\alpha$ cluster distributions
in both the $0^+$ and first excited $2^+$ states.
The $\rho_p(k)$ for the $2^+$ state at 3 MeV excitation, which is not shown, 
is almost identical
with the ground state, varying by $+3$ to $-5\%$ out to 5 fm$^{-1}$.
At very low momenta, the nucleus can be considered to be primarily
in an $\alpha\alpha$ cluster, but this probability declines rapidly as $k$ 
increases.
The $\rho_{\alpha\alpha}(k)$ for the $0^+$ state has two nodes because the 
$\alpha$'s are in a relative $2s$ spatial state, but for the $2^+$ rotational
excited state they are in a relative $1d$ spatial state.
The integrated $N_{\alpha\alpha}$ are 0.84 and 0.83, respectively.
The next state in the $^8$Be spectrum is the $4^+$ rotational state at 11 MeV
excitation; its $\rho_p(k)$ varies from the ground state by $+7$ to $-14\%$
over the momentum range shown here, while its $\rho_{\alpha\alpha}(k)$
exhibits the structure of a $0g$ spatial state, with a smaller
$N_{\alpha\alpha} = 0.75$.
The $\rho_p(k)$ and $\rho_{\alpha\alpha}(k)$ for all three states are
tabulated online.

The proton and neutron momentum distributions in $^{10}$Be are shown
in Fig.~\ref{fig:be10}.
The proton distribution is very much like that in $^8$Be, while the
two added neutrons are mostly at small, but finite $k$ as appropriate
for the $p$-shell.
Also shown is the difference $\rho_n(k)-\rho_p(k)$; this difference
is very much like the two halo neutrons in $^6$He shown in Fig.~\ref{fig:he6},
again with a dip at zero momentum, 
and with a dip at the usual $S$-wave node position, although it is somewhat
smeared out compared to the former case.

Tables and figures of the nucleon momentum distributions are provided online at
http://www.phy.anl.gov/theory/research/momenta/.
The tables give the proton momentum distribution, with one standard deviation
error bars, in $T$=0 nuclei with the understanding that neutron and
proton momentum distributions are identical.
For $T>0$ nuclei, both proton and neutron momentum distributions are 
tabulated.
Similarly, for $J \!=\! 0$ nuclei, only total nucleon momentum distributions
are given, but for $J>0$, the spin-up and spin-down distributions for
$M_J \!=\! J$ states are also given.
In addition, the total normalization $N_{\sigma\tau}$ for each momentum 
distribution is given, as are the contributions to the kinetic energy.
The corresponding spatial densities for these wave functions are provided 
online at http://www.phy.anl.gov/theory/research/density/.

Included in the tables, but not shown here, are distributions for the ground
states of $^3$H, $^8$He, $^{8,9}$Li, $^9$Be, and $^{10}$B.
We have also made preliminary calculations of the nucleon momentum 
distributions in $^{11}$B and $^{12}$C.
In these cases, the Jastrow part of the variational wave function has been 
made with only the most spatially symmetric components, [443] and [444],
respectively.
The web site will be updated when calculations are made with
more sophisticated wave functions for these and other nuclei.

\begin{figure}[t!]
\includegraphics[angle=-90,width=3.25in]{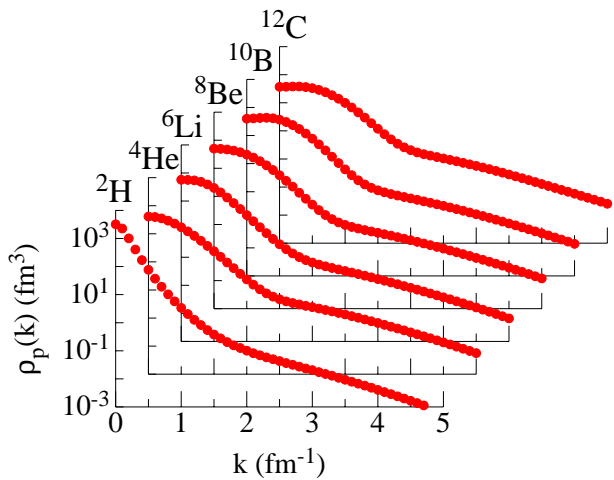}
\caption{(Color online) The proton momentum distributions in 
all $T$=0 nuclei from $A$=2--12.}
\label{fig:allt0}
\end{figure}

The overall evolution of the proton momentum distribution in the $T$=0
nuclei is illustrated in Fig.~\ref{fig:allt0}.
The shape of the distributions shows a smooth progression as nucleons
are added.
As $A$ increases, the nuclei become more tightly bound, and the fraction
of nucleons at zero momentum decreases.
As nucleons are added to the $p$-shell, the distribution at low momenta
becomes broader, and develops a peak at finite $k$.
The sharp change in slope near $k=2$ fm$^{-1}$ to a broad shoulder is 
present in all these nuclei and is attributable to the strong tensor
correlation induced by the pion-exchange part of the $N\!N$ potential,
boosted by the two-pion-exchange part of the $3N$ potential.
Above $k=4$ fm$^{-1}$, the bulk of the momentum density appears to come from
short-range spin-isospin correlations.

\section {Nucleon-pair results}
\label{sec:double}

The probability of finding two nucleons in a nucleus with relative momentum
${\bf q}=({\bf k}_1-{\bf k}_2)/2$ and total center-of-mass momentum
${\bf Q}={\bf k}_1+{\bf k}_2$ in a given spin-isospin state is given by:
\begin{eqnarray}
\label{eq:rhoqQ}
\rho_{ST}({\bf q},{\bf Q}) &=&
\int d{\bf r}^\prime_1 d{\bf r}_1 d{\bf r}^\prime_2d{\bf r}_2 d{\bf r}_3 \cdots d{\bf r}_A \nonumber \\
&& \psi^\dagger_{JM_J}({\bf r}_1^\prime,{\bf r}^\prime_2,{\bf r}_3,\dots,{\bf r}_A)  \\
&& e^{-i{\bf q}\cdot({\bf r}_{12}-{\bf r}_{12}^\prime)} 
e^{-i{\bf Q}\cdot({\bf R}_{12}-{\bf R}_{12}^\prime)} \nonumber \\
&&  P_{ST}(12) \psi_{JM_J}({\bf r}_1,{\bf r}_2,{\bf r}_3,\dots,{\bf r}_A) \, ,
\nonumber
\end{eqnarray}
where ${\bf r}_{12} = {\bf r}_1 - {\bf r}_2$,
${\bf R}_{12} = ({\bf r}_1 + {\bf r}_2)/2$,
and $P_{ST}(12)$ is a projector onto pair spin $S=0$ or 1, and isospin
$T=0$ or 1.
The total normalization is:
\begin{equation}
N_{ST} = \int \frac{d{\bf q}}{(2\pi)^3} \frac{d{\bf Q}}{(2\pi)^3}
         \rho_{ST}({\bf q},{\bf Q}) \, ,
\end{equation}
where $N_{ST}$ is the total number of nucleon pairs with given spin-isospin..
Alternate projectors can also be used, e.g., for $N\!N$ pairs $pp$, $np$, 
and $nn$ with corresponding normalizations.

\begin{figure}[t!]
\includegraphics[angle=-90,width=3.25in]{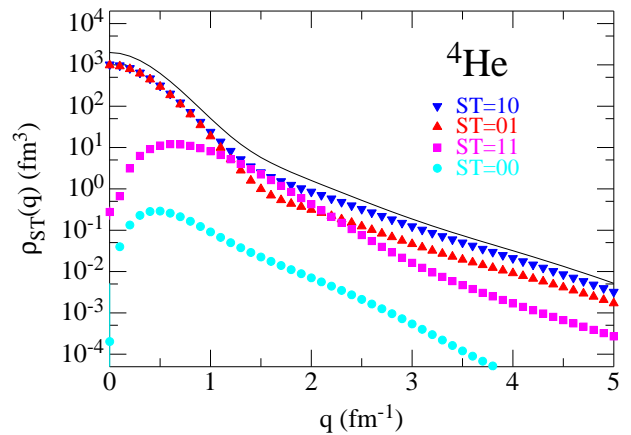}
\caption{(Color online) The nucleon-pair momentum distributions in $^4$He 
as a function of relative momentum $q$ is shown by the solid black line,
while the projection into pairs of total spin-isospin $ST$ is given by
the various symbols.}
\label{fig:he4st}
\end{figure}

The nucleon-pair momentum distributions can be examined in a number of
different ways.
One way is to integrate over all values of ${\bf Q}$ and reduce the pair
density to a function $\rho_{12}(q)$ of the relative momentum $q$ only. 
In this case, Eq.(\ref{eq:rhoqQ}) reduces to a form similar to
Eq.(\ref{eq:actual}), with a sum over all configurations in the Monte
Carlo walk controlled by $|\Psi_{JM_J}|^2$, and a Gauss-Legendre integration 
over the relative separation ${\bf x} = {\bf r}_{12}-{\bf r}_{12}^\prime$.
Again, the polar angle ${\bf \Omega}_x$ is sampled by randomly choosing the
direction of ${\bf x}$ in space, and an average over all pairs in every
MC configuration is made.

\begin{figure}[b!]
\includegraphics[angle=-90,width=3.25in]{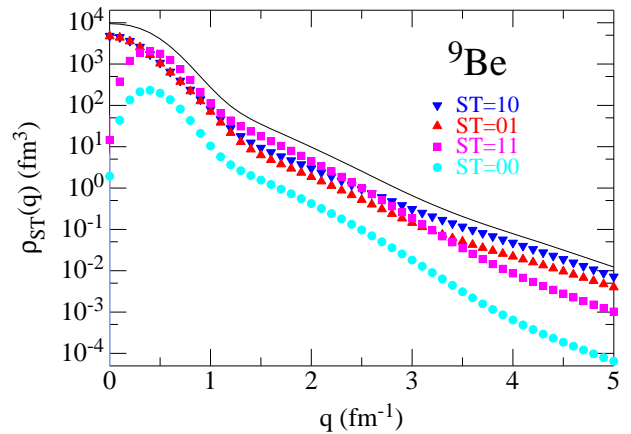}
\caption{(Color online) The nucleon-pair momentum distributions in $^9$Be 
as a function of relative momentum $q$ is shown by the solid black line,
while the projection into pairs of total spin-isospin $ST$ is given by
the various symbols.}
\label{fig:be9st}
\end{figure}

Typical results are shown in Fig.~\ref{fig:he4st} for $^4$He and in 
Fig.~\ref{fig:be9st} for $^9$Be, where both the total distribution and the 
different $ST$ components are plotted.
The spatial distribution of nucleon pairs in different $ST$ combinations,
using correlated wave functions generated by realistic interactions, was
first studied in Ref.~\cite{FPPWSA96} for $A \leq 7$ nuclei.
The corresponding momentum distributions have been calculated here for
a number of cases in addition to those shown in the figures, including $^3$He, 
$^{6,7}$Li, $^8$Be, $^{10}$B, and are available online~\cite{momenta}.

\begin{table}[t!]
\caption{Total number of spin-isospin $ST$ pairs in different nuclei using
both independent pair (IP) and correlated (cor) wave functions.}
\begin{ruledtabular}
\begin{tabular}{ l c d d d d }
Nucleus & $\Psi$ & \multicolumn{1}{c}{$N_{01}$} 
                 & \multicolumn{1}{c}{$N_{11}$}
                 & \multicolumn{1}{c}{$N_{10}$}
                 & \multicolumn{1}{c}{$N_{00}$} \\
\hline
$^3$He($\frac{1}{2}^+$)  & IP  & 1.5   & 0.0   & 1.5   & 0.0   \\
                         & cor & 1.36  & 0.14  & 1.49  & 0.01  \\
$^4$He($0^+$)            & IP  & 3.0   & 0.0   & 3.0   & 0.0   \\
                         & cor & 2.54  & 0.46  & 2.99  & 0.01  \\
                   & [AV18)    & 2.60  & 0.40  & 2.99  & 0.01  \\
                   & [AV18(4)] & 2.99  & 0.01  & 2.99  & 0.01  \\
                   & [AV18(J)] & 3.00  & 0.00  & 3.00  & 0.00  \\
$^6$Li($1^+$)            & IP  & 4.5   & 4.5   & 5.5   & 0.5   \\
                         & cor & 4.05  & 4.95  & 5.47  & 0.53  \\
$^7$Li($\frac{3}{2}^-$)  & IP  & 6.75  & 6.75  & 6.75  & 0.75  \\
                         & cor & 6.12  & 7.38  & 6.74  & 0.76  \\
$^8$Be($0^+$)            & IP  & 9.0   & 9.0   & 9.0   & 1.0   \\
                         & cor & 8.07  & 9.93  & 8.98  & 1.02  \\
$^9$Be($\frac{3}{2}^-$)  & IP  & 10.5  & 13.5  & 10.5  & 1.5   \\
                         & cor & 9.56  & 14.44 & 10.48 & 1.52  \\  
$^{10}$B($3^+$)          & IP  & 12.0  & 18.0  & 13.0  & 2.0   \\
                         & cor & 11.17 & 18.83 & 12.95 & 2.05
\label{tab:ST}
\end{tabular}
\end{ruledtabular}
\end{table}

The number of $N_{ST}$ pairs for these nuclei are given in Table~\ref{tab:ST}.
In the absence of spin-isospin correlations, the total number of $ST$ pairs in
a given nucleus depends on the total number of $N\!N$ pairs
$P_A = A(A-1)/2$, the total nuclear spin $S_A$ and isospin $T_A$, and the 
spatial symmetry of the wave function as described in Ref.~\cite{W06}. 
This is the independent-pair (IP) value given in the table.
(For $p$-shell nuclei there are multiple spatial symmetry states available;
we quote the IP values for the most symmetric state, which is always the
dominant ground-state component for realistic $N\!N$ interactions.)
For the $s$-shell nuclei, only $ST$ combinations (10) and (01) with even 
orbital angular momentum are present.
In the $p$-shell, the $ST$ combinations (11) and (00) with odd
orbital angular momentum start to contribute. 
For $A \geq 9$ nuclei, the (11) pairs become the most numerous of all;
in the nuclear matter limit, the ratio (11):(10):(01):(00) is 9:3:3:1.

When spin-isospin correlations, and particularly the tensor-isospin 
correlations, are switched on, there is a noticeable transmutation of 
$ST$ = (01) pairs into (11) pairs, relative to the IP values.
There is also a much smaller conversion of $ST$ = (10) pairs to (00) pairs.
(Because isospin is largely conserved for realistic nuclear forces,
we use wave functions of fixed total isospin, so the sum of (01) and (11) 
pairs is unchanged from the IP value, as is the sum of (10) and (00) pairs.)
These correlated values are in the rows labeled "cor" in Table~\ref{tab:ST}.
This transmutation comes about because tensor correlations can flip the spin
of a nucleon~\cite{FPPWSA96}, e.g., if a pair of nucleons is in a (01) 
combination, a tensor correlation to a third nucleon has an amplitude to 
flip the spin of one of the pair nucleons and change the pair to (11).
Because realistic $N\!N$ interactions are much more repulsive in $ST$ = (00)
than (11) combinations, optimized correlations will only permit a much smaller
transmutation of (10) pairs into (00) pairs.

That the tensor correlations are the source of this effect is demonstrated in 
the case of $^4$He by calculations using the AV18 interaction alone, and the 
AV18(4) and AV18(J) truncated wave functions discussed above in conjunction
with Fig.~\ref{fig:he4};
their $N_{ST}$ values are given in Table~\ref{tab:ST} in the rows labeled
[AV18], [AV18(4)], and [AV18(J)].
The pure Jastrow wave function without any spin-isospin correlations generates
no (11) or (00) pairs at all.
The AV18(4) wave function without tensor forces generates
(11) and (00) pairs with only a tiny amplitude of 0.01.
The AV18 interaction with a realistic tensor force converts 0.40 (01) 
pairs to (11) pairs. 
Addition of the $3N$ potential boosts this effect about 15\% to 0.46 pairs,
proportional to the increase it induces in the $^4$He $D$-state percentage.
From the results shown in Table~\ref{tab:ST} for $^4$He and $^8$Be,
from our preliminary results for $^{12}$C, and from the $^{16}$O and $^{40}$Ca
results of Ref.~\cite{Alvioli13}, it appears that $\sim$0.5 (10) pairs 
are converted to (11) pairs for every $\alpha$-particle cluster in the nucleus.

In the $\rho_{12}(q)$ projections for $^4$He shown in Fig.~\ref{fig:he4st}, 
the even partial-wave $ST$ = (10) and (01) pairs are seen to have a similar 
shape, with an $S$-wave peak out to $q$=1 fm$^{-1}$ followed by a
change of slope with a broad shoulder extending beyond 5 fm$^{-1}$.
At higher momenta, the (10) pairs dominate, while the (01) pairs show
a relative dip in the $q$=1.5--2 fm$^{-1}$ range.
The odd partial-wave (predominantly $P$-wave) (11) pairs vanish at $q=0$
but grow to equal the (10) pairs and exceed the (01) pairs in the dip region.
The (00) pairs also have a $P$-wave shape, but are much smaller in magnitude.
For $q \geq 1$ fm$^{-1}$ there are 0.56 (10) pairs compared to 
0.29 (01) pairs and 0.29 (11) pairs.

The $\rho_{12}(q)$ projections for $^9$Be shown in Fig.~\ref{fig:be9st} have a 
very similar structure, except that the total number of (11) pairs has grown 
significantly (as expected from the IP count) to become the largest component
for $q$=0.5--2.5 fm$^{-1}$.
The number of (00) pairs has also grown, but remains much smaller than
the others.
At higher momenta, the (10) pairs still dominate the tail of the momentum
distribution.

\begin{figure}[t!]
\includegraphics[angle=-90,width=3.25in]{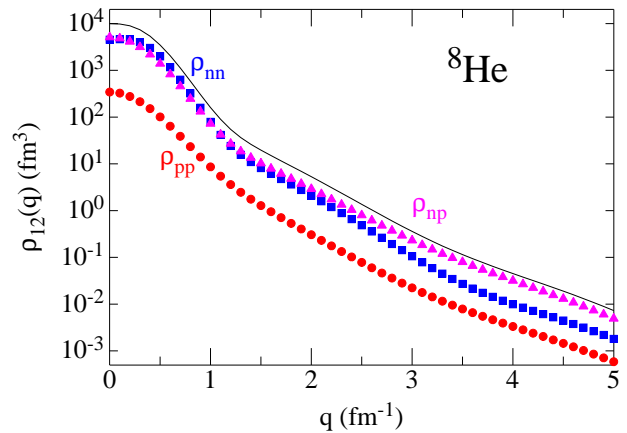}
\caption{(Color online) The nucleon-pair momentum distributions in $^8$He 
as a function of relative momentum $q$ is shown by the solid black line,
while the projection into $nn$, $np$, and $pp$ pairs is shown by the
blue squares, magenta triangles, and red circles, respectively.}
\label{fig:he8qNN}
\end{figure}

Instead of projecting $\rho_{12}(q)$ into $ST$ pairs, we can
project out $pp$, $np$, and $nn$ pairs.  An example of this 
is shown in Fig.~\ref{fig:he8qNN} for $^8$He.
In this neutron-rich nucleus, there are 15 $nn$ pairs, 12 $np$ pairs,
and only 1 $pp$ pair.
The excess of $nn$ over $np$ pairs is all at low relative momenta
$q \leq 1$ fm$^{-1}$, where 13.5 (10.2) $nn$ ($np$) pairs reside.
The total number of pairs with $q \geq 2$ fm$^{-1}$ is small, but
there are 70\% more $np$ than $nn$ pairs in this regime.
Calculations of $\rho_{N\!N}(q)$ for $^{4,6}$He, $^6$Li, $^8$Be, and $^{10}$B 
are also provided online~\cite{momenta}.

The $\rho_{pp}(k)$ in the helium isotopes characterizes the $s$-shell core
and shows only small changes as neutrons are added to go from $^4$He to $^6$He 
and $^8$He.
This is consistent with the findings of Ref.~\cite{WPCP00} where only
small differences in the two-proton spatial densities were found, indicating
the $\alpha$ core in $^{6,8}$He is fairly stiff against deformation.

\begin{figure}[t!]
\includegraphics[angle=-90,width=3.25in]{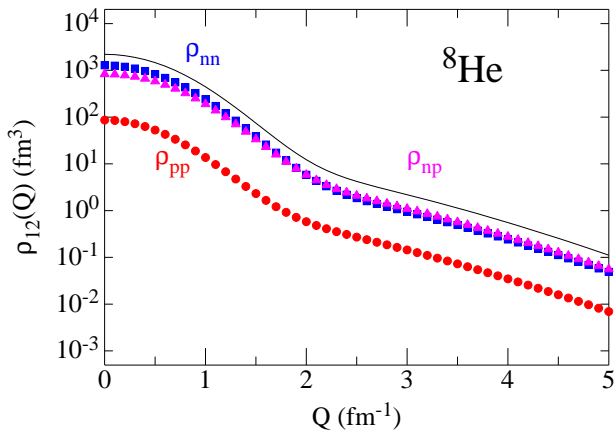}
\caption{(Color online) The nucleon-pair momentum distributions in $^8$He 
as a function of total pair momentum $Q$ is shown by the solid black line,
while the projection into $nn$, $np$, and $pp$ pairs is shown by the
blue squares, magenta triangles, and red circles, respectively.}
\label{fig:he8QNN}
\end{figure}

We can also integrate Eq.(\ref{eq:rhoqQ}) over all ${\bf q}$,
leaving a function $\rho_{12}(Q)$ of the total pair momentum $Q$ only.
In Fig.~\ref{fig:he8QNN} we show $\rho_{12}(Q)$ projected into $N\!N$ pairs,
again for $^8$He.
In general, the $\rho_{N\!N}(Q)$ for a given nucleus has a smaller falloff 
at large momenta than the $\rho_{N\!N}(q)$ and the ratios of different 
$N\!N$ components vary less over the range of $Q$.
In addition to $^8$He, calculations of $\rho_{N\!N}(Q)$ for $^{3,4,6}$He, 
$^6$Li, and $^8$Be are provided online~\cite{momenta}.

A simple internal consistency check on our integration methods
can be made using $^3$He.
By virtue of momentum conservation, when a $pp$ pair has total momentum 
${\bf Q}$, the remaining neutron has momentum ${\bf k} = -{\bf Q}$.
Consequently, $\rho_{pp}(Q) = \rho_n(k)$ for $k = Q$ and similarly
$\rho_{np}(Q) = \rho_p(k)$.
This relation is satisfied to better than 1\% as seen by comparing
the relevant online tables.
We can also again contrast the $\rho_{pp}(Q)$ among the $^{4,6,8}$He
isotopes; they are similar in shape, but show a noticeably greater
variation than the $\rho_{pp}(q)$ because the center-of-mass is no
longer at the center of the $\alpha$ core once the $p$-shell neutrons are added.
The $\rho_{pp}(Q=0)$ are ordered proportional to the charge radius,
i.e., the value for $^6$He is largest and that for $^4$He is smallest.

\begin{figure}[t!]
\includegraphics[angle=-90,width=3.25in]{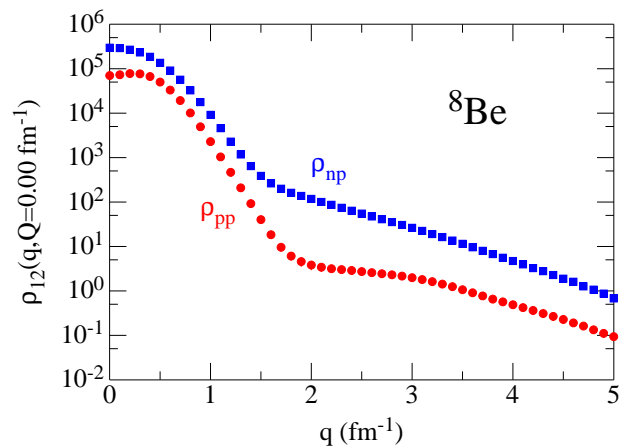}
\caption{(Color online) The nucleon-pair momentum distributions in $^8$Be 
for back-to-back $Q$=0 pairs as a function of relative momentum $q$ is 
shown by red circles for $pp$ pairs and blue squares for $np$ pairs.}
\label{fig:be8qQ}
\end{figure}

Nucleon-pair momentum distributions $\rho_{12}({\bf q},{\bf Q})$ as a 
function of both ${\bf q}$ and ${\bf Q}$ are more expensive to evaluate as 
Eq.(\ref{eq:rhoqQ}) then requires a double Gauss-Legendre integration for
each pair in each MC configuration; see Eq.(3) of Ref.~\cite{SWPC07}.
Nucleon-pair distributions for the special case $Q$=0 were evaluated 
for a number of $A \leq 8$ nuclei in~\cite{SWPC07}, to help understand the
results of a two-nucleon knockout experiment on $^{12}$C at Jefferson
Lab~\cite{Subedi08}.
The experiment showed a ratio of back-to-back $np$ to $pp$ pairs of nearly 20
(compared to the simple total ratio of 2.4) in the relative momentum range 
$q$=1.5--3 fm$^{-1}$.
This is easily understood by the presence of a noticeable dip in the $Q$=0
$pp$ distribution for this range of $q$ relative to the $np$ distribution,
as shown in Fig.~\ref{fig:be8qQ} for $^8$Be.
In addition to this case, our latest results at $Q$=0 for $^{3,4}$He, 
$^{6,7}$Li, and $^{10}$B are provided in the online tables~\cite{momenta}.
Unfortunately, $^{12}$C is currently still too expensive to evaluate
with this double integral, but we are exploring more efficient ways of
evaluating Eq.(\ref{eq:rhoqQ}) that should make it possible in the future.

\begin{figure}[t!]
\includegraphics[angle=-90,width=3.25in]{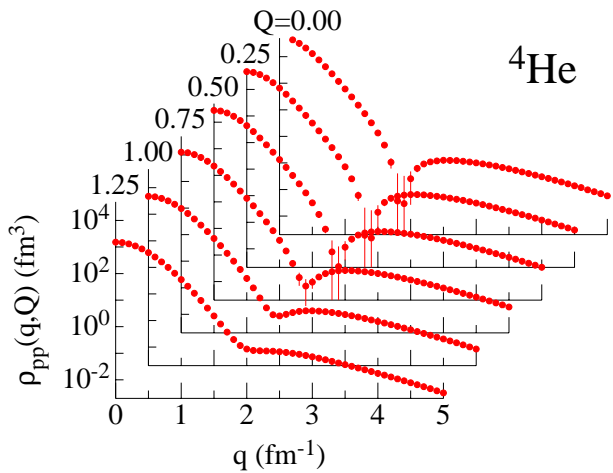}
\caption{(Color online) The proton-proton momentum distributions in $^4$He 
averaged over the directions of ${\bf q}$ and ${\bf Q}$ as a function of $q$ for several fixed values of $Q$ from 0 to 1.25 fm$^{-1}$.}
\label{fig:pphe4}
\end{figure}

\begin{figure}[t!]
\includegraphics[angle=-90,width=3.25in]{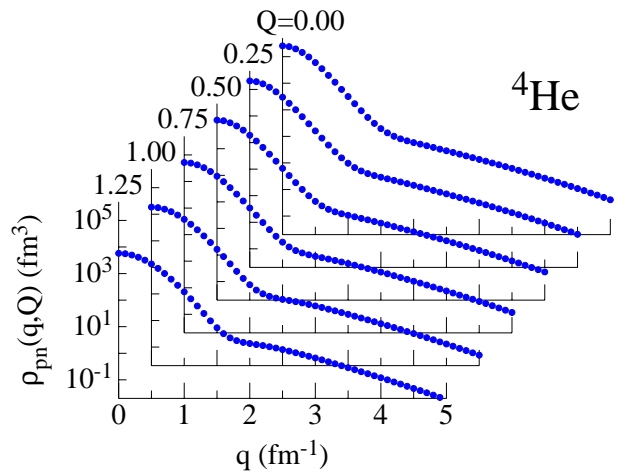}
\caption{(Color online) The proton-neutron momentum distributions in $^4$He 
averaged over the directions of ${\bf q}$ and ${\bf Q}$ as a function of $q$ for several fixed values of $Q$ from 0 to 1.25 fm$^{-1}$.}
\label{fig:pnhe4}
\end{figure}

Nucleon-pair distributions for $^3$He and $^4$He with finite 
${\bf Q} \parallel {\bf q}$ were reported in~\cite{WSPC08}.
We now have made much more extensive calculations for these two nuclei
averaged over all directions of ${\bf Q}$ and ${\bf q}$.
The $\rho_{pp}(q,Q)$ for $^4$He is shown in Fig.~\ref{fig:pphe4}
and the $\rho_{pn}(q,Q)$ is shown in Fig.~\ref{fig:pnhe4}
as a series of $q$-dependent curves for fixed values of $Q$ from 0 to 
1.25 fm$^{-1}$.

The trends illustrated by these figures are similar to the single-nucleon
momentum distributions.
The $pp$ pairs are primarily in relative $^1$S$_0$ states, and exhibit
the usual $S$-wave node for total pair momenta $Q$=0.
As $Q$ increases, this node begins to fill in, until it is only a dip
for $Q\!>\!1$ fm$^{-1}$.
The $pn$ pairs are predominantly in deuteron-like states, with the
$D$-wave contribution filling in the $S$-wave node and beginning the usual
broad shoulder above 2 fm$^{-1}$.
The overall magnitude of the curves for both $pp$ and $pn$ pairs decreases
as $Q$ increases, simply because there are fewer pairs with high
total momenta.
The numerical values for these curves may be found online along with similar
results for $^3$He~\cite{momenta}.
Additional calculations for various fixed angles between ${\bf Q}$ and
${\bf q}$ are in progress.

\section {Conclusions}
\label{sec:conclude}

We have calculated a large number of nucleon, nucleon-cluster, and
nucleon-pair momentum distributions for $A \leq 12$ nuclei using
VMC wave functions generated from a realistic Hamiltonian containing 
the AV18 $N\!N$ and UX $3N$ potentials.
The single-nucleon $\rho(k)$ have a common characteristic shape,
with a peak at zero (for $s$-shell nuclei) or small (for $p$-shell nuclei)
momentum, a rapid drop of more than two orders of magnitude
followed by a distinct change of slope around $q$=1.5--2 fm$^{-1}$,
and then an extended high-momentum tail that carries well beyond 5 fm$^{-1}$.
The dominant source of this tail in the 1.5--3 fm$^{-1}$ range is the 
$N\!N$ tensor force, which comes from the pion-exchange potential, and
is boosted $\sim$15\% by the $3N$ force with its two-pion-exchange terms.
Spin-isospin correlations appear to dominate at higher momenta, and 
presumably are dependent on the short-range structure of the Hamiltonian.
Our calculations include a breakdown between spin-up and spin-down
nucleons in $J>0$ nuclei and between protons and neutrons in $T>0$ nuclei.
Nucleon-cluster distributions, with $d$, $t$, and $\alpha$ clusters, 
have been calculated in $A \leq 8$ nuclei, including a number of excited states.
They do not exhibit high-momentum tails but have specific nodal structures
that reflect the requirements of antisymmetry for the given spin $J$
of the nuclear state.

The nucleon-pair momentum distributions $\rho_{12}(q)$ as a function
of relative pair momentum $q$ have a very similar structure to $\rho(k)$.
We provide the breakdown into pair spin-isospin $ST$ channels in a 
number of cases, and alternatively into $N\!N$ channels in other cases.
These different components have different behaviors as a function of $q$
which can again be traced back primarily to the pion-exchange tensor force,
with the high-momentum tails largely dominated by the $ST$ = (10) or $np$ pairs.
The distributions $\rho_{12}(Q)$ as a function of total pair momentum $Q$
also have a similar shape, but are noticeably flatter and seem to have
less interesting structure.
The $\rho_{12}({\bf q},{\bf Q})$ are more complicated, particularly with
the evolution of nodal structures as illustrated in Fig.\ref{fig:pphe4}.

While we have illustrated many of the above cases in the figures of this
paper, the main results are to be found in our online figures and
tabulations~\cite{momenta}.
These include single-nucleon momentum distributions for 16 different nuclear
ground states plus some excited states.
The corresponding configuration-space densities are also provided.
Nucleon-cluster distributions are given for 12 different cases.
Nucleon-pair momentum distributions include eight $\rho_{ST}(q)$,
seven $\rho_{N\!N}(q)$, six $\rho_{N\!N}(Q)$, and six $\rho_{N\!N}(q,Q=0)$
nuclear ground states, plus two dozen $\rho_{N\!N}(q,Q>0)$ cases
in $^{3,4}$He.
It is our intention to add additional calculations as they are developed and
requests for specific cases will be entertained.

In the future it may be possible to evaluate the momentum distributions
using the more accurate GFMC wave functions~\cite{BPW11}.
In that case, Eq.(\ref{eq:rhok}) will have to be evaluated moving only 
${\bf r}^\prime_i$, which is an inherently noisier procedure.
The most likely result from a GFMC evaluation is a further shift to
higher momenta.
First, the GFMC wave functions produce more binding which will decrease
somewhat the population of nucleons at low momenta.
Second, GFMC wave functions generally have spin-isospin and tensor correlations
that are enhanced relative to the VMC wave functions, which will
raise the broad shoulder in the 2 to 7 fm$^{-1}$ region;
these correlations should also enhance polarization.
The GFMC kinetic energy, i.e., the second moment of the nucleon
momentum distribution, is generally higher than the VMC kinetic energy
by $\sim 10\%$, also indicating a shift of nucleons to higher momenta,
but still subject to the overall normalization remaining the same.

\section*{Acknowledgments}

This work is supported by the U.S. Department of Energy, Office of
Nuclear Physics, under contracts DE-AC02-06CH11357 (RBW and SCP),
DE-AC05-06OR23177 (RS), DE-AC52-06NA25396 (JC), and the NUCLEI SciDAC grant.
The calculations were made on the parallel computers of Argonne's Laboratory 
Computing Resource Center.
\end{document}